\documentclass[letterpaper]{article}
\usepackage{aaai24}
\usepackage{times}
\usepackage{helvet}
\usepackage{courier}
\usepackage[hyphens]{url}
\usepackage{graphicx}
\usepackage{natbib}
\usepackage{caption}
\usepackage{algorithm}
\usepackage{algorithmic}

\usepackage{newfloat}
\usepackage{listings}
\DeclareCaptionStyle{ruled}{labelfont=normalfont,labelsep=colon,strut=off}
\floatstyle{ruled}
\newfloat{listing}{tb}{lst}{}
\floatname{listing}{Listing}
\usepackage{caption}
\usepackage{subcaption}
\usepackage{xcolor}
\usepackage{xspace}
\usepackage{booktabs}

\usepackage{colortbl}
\usepackage{enumitem}
\usepackage{multirow}
\usepackage{pifont}
\usepackage{tabularx}

\title{
Preventing Eviction-Caused Homelessness through ML-Informed Distribution of Rental Assistance
}
\author{
Catalina Vajiac\equalcontrib\textsuperscript{\rm 1},
Arun Frey\equalcontrib\textsuperscript{\rm 2},
Joachim Baumann\equalcontrib\textsuperscript{\rm 3,\rm 4},
Abigail Smith\equalcontrib\textsuperscript{\rm 5},\\
Kasun Amarasinghe\textsuperscript{\rm 1},
Alice Lai\textsuperscript{\rm 1},
Kit Rodolfa\textsuperscript{\rm 2},
Rayid Ghani\textsuperscript{\rm 1}
}
\affiliations{
\textsuperscript{\rm 1}Carnegie Mellon University,\\
\textsuperscript{\rm 2}Stanford University,\\
\textsuperscript{\rm 3}University of Zurich,\\
\textsuperscript{\rm 4}Zurich University of Applied Sciences,\\
\textsuperscript{\rm 5}NORC at the University of Chicago\\
cvajiac@cs.cmu.edu,
arunfrey@law.stanford.edu,
baumann@ifi.uzh.ch,
als1@u.northwestern.edu,
\{kamarasi, alicelai, ghani\}@andrew.cmu.edu,
krodolfa@law.stanford.edu
}

\newcommand{\pspace}{\hspace{.5cm}}
\newcommand{\model}[1]{\textbf{#1}}
\newcommand{\param}[1]{\pspace\textit{#1}}

\newcommand{\eat}[1]{}

\newcommand{\precision}{precision@100\xspace}

\newcommand{\blhomeless}{\textit{B1}\xspace}
\newcommand{\blbaserate}{\textit{B2}\xspace}
\newcommand{\blofp}{\textit{B3}\xspace}
\newcommand{\blhomelessfull}{\textit{{\blhomeless: Previously Homeless}}\xspace}
\newcommand{\blbaseratefull}{\textit{{\blbaserate: Baserate}}\xspace}
\newcommand{\blofpfull}{\textit{{\blofp: Early OFP}}\xspace}
\newcommand{\missedgroup}{\textit{missed group}\xspace}

\definecolor{lightblue}{HTML}{cee4f0}

\newcommand{\codeurl}{\url{https://github.com/dssg/acdhs_housing_public}}
\newcommand{\Allegheny}{Allegheny\xspace}
\newcommand{\AlleghenyCounty}{Allegheny County\xspace}
\newcommand{\DHS}{Department of Human Services\xspace}
\newcommand{\ACDHS}{ACDHS\xspace}
\newcommand{\ACDHSs}{ACDHS'\xspace}

\newlength{\wdth}

\newcommand{\tablebullet}{\ding{226}\xspace}

\begin{document}

\maketitle

\begin{abstract}
Rental assistance programs provide individuals with financial assistance to prevent housing instabilities caused by evictions and avert homelessness. Since these programs operate under resource constraints, they must decide who to prioritize. Typically, funding is distributed by a reactive or first-come-first serve allocation process that does not systematically consider risk of future homelessness. We partnered with \Allegheny County, PA to explore a proactive allocation approach that prioritizes individuals facing eviction based on their risk of future homelessness. Our ML system that uses state and county administrative data to accurately identify individuals in need of support outperforms simpler prioritization approaches by at least 20\% while being fair and equitable across race and gender. Furthermore, our approach would identify 28\% of individuals who are overlooked by the current process and end up homeless. Beyond improvements to the rental assistance program in \Allegheny County, this study can inform the development of evidence-based decision support tools in similar contexts, including lessons about data needs, model design, evaluation, and field validation.
\end{abstract}

\section{Introduction}

Homelessness remains a pervasive and pressing issue across the United States.
In January 2022, more than 500k individuals experienced homelessness on a single night \cite{desousa_2022}. Rising eviction rates and the lack of affordable housing contribute to this problem: as the gap between housing costs and income levels continues to grow, an increasing number of households struggle to pay rent, eventually facing eviction and, in some cases, homelessness.

To curb rates of homelessness, policymakers are increasingly seeking to reduce the rate of entry into homelessness \cite{culhane_2011}. One popular prevention strategy is rental assistance programs, which provide temporary financial assistance to individuals facing eviction to keep them stably housed. Experimental evidence suggests that such assistance is effective at reducing subsequent homelessness:
using quasi-random variation in funding availability of a rental assistance program in Chicago, it was shown that individuals who called when funding was available were 76\% less likely to become homeless in the subsequent 6 months compared to those who called when funding was unavailable~\cite{evans_2016}.
A similar program in NYC randomized the provision of financial assistance to households and reduced average days in shelter from 32 nights among the control group to 10 nights among the treatment group in the 2 years following funding \cite{rolston_2013}.

For rental assistance programs to be a viable homelessness prevention strategy, they need to be effective (i.e. prevent individuals from falling into homelessness), efficient (i.e. target individuals at risk of falling into homelessness) \cite{burt_2007}, and equitable. 
Currently, however, these programs prioritize individuals based on simple heuristics (e.g. first come, first served) instead of their likelihood of future homelessness. In addition, they place the burden of applying for funding on the individuals facing eviction and overlook those who need help but do not apply. The administrative process can create long delays, so that even eligible tenants often remain on a waitlist for multiple months and can end up evicted before they receive assistance. 

In this paper, we describe our collaboration with the \Allegheny County \DHS (\ACDHS) to improve the allocation of rental assistance by prioritizing individuals with the highest likelihood of falling into homelessness. We combine rich county and state administrative data with court records to predict homelessness among all individuals currently facing eviction, regardless of whether they contacted the county for financial support. Using data from January 2012 through August 2023, we develop a series of machine learning (ML) models that predict a tenant’s need for homelessness services in the next 12 months, allowing \ACDHS to proactively prioritize rental assistance to the most vulnerable tenants. In particular, we make the following contributions:

\begin{itemize}
    \item \textbf{Enabling a proactive approach:}   
    We consider a significantly higher proportion of residents, namely all tenants in the county facing eviction, instead of only those who call the county for help. Our models identify 28\% of people who are overlooked by the current process and end up homeless. By shifting the burden away from those impacted, this proactive approach also reduces administrative effort and is more likely to provide preventative rental assistance to tenants before their eviction.
    \item \textbf{Implementing need-based prioritization:} Our models identify individuals in need of future homelessness support services with at least 20\% improvement over simpler baselines, and are 10x better than random selection while also being fair and equitable.
    \item \textbf{Field validation:} We conduct a shadow mode deployment to validate our models on new data, mitigating the risk of leakage in the future. Additionally, we are planning a randomized control trial to compare our proposed solution to the current process and to evaluate the effectiveness of rental assistance in preventing entry into homelessness among targeted individuals. 
    \item \textbf{Lessons learned:} We reflect on pitfalls and successes that may inform AI researchers seeking to ethically design predictive decision support tools in other contexts.
\end{itemize}

\paragraph{Reproducibility.} Our code is available at \codeurl.

\section{Ethical Considerations}

Since we are informing the allocation of scarce and critical resources to vulnerable people who could become homeless, we bring into focus the ethical considerations that informed and were embedded in every phase of the scoping, design, and development of our approach.  

\paragraph{Equitable outcomes through the allocation of resources.} The use of AI in real-world contexts can perpetuate systemic biases such as racial disparities~\cite{chen2020treating}.
To mitigate this risk, we carefully designed the scope of our work and our formulation and analyzed our model results to guard against bias against certain demographic groups, i.e. race and gender (see Section~\ref{ssec:results-fairness}). Our field trial will also test the equity in the impact of the downstream decisions made using our system.

\paragraph{Transparency and interpretability.} Our goal is to develop a system that supports and informs social workers in making better decisions that lead to improved, more equitable outcomes.
To this end, we train models that are more explicit about which features they learn from (see Section~\ref{ssec:model-training}) and analyze which features end up most predictive of future homelessness (see Section~\ref{ssec:feature-importance}). Our solution is designed to augment the current process, supporting rather than excluding social workers from the decision-making loop and providing them with additional information beyond predictions to help them make better-informed decisions. We also consulted impacted community members in the design and validation process.

\paragraph{Privacy concerns of using sensitive data.} The collection of personal information always comes with privacy concerns.
In our case, particularly because \ACDHS connects administrative data across different facets of residents' lives, we want to be sure that the use of this data offers significant enough improvements to the community, particularly people potentially facing homelessness. In Sections~\ref{sec:key-findings} and \ref{sec:validation}, we extensively evaluate our models with historical data, and a field trial to ensure that this work improves outcomes. We are currently soliciting feedback from groups potentially impacted by the system through a community engagement process to better assess tradeoffs before conducting a validation trial.

\section{Current Approaches}
We first consider how \ACDHS currently prioritizes residents for rental assistance and highlight relevant prior research.

\subsection{\ACDHSs Current Process}
The current process for obtaining rental assistance in \Allegheny County is illustrative of how such programs are usually managed throughout the United States. Tenants with an eviction notice can contact the \AlleghenyCounty Link helpline to request rental assistance, or they may be referred to the rental assistance program by a mediation program or a housing assistance organization. Applicants are pre-screened according to available funding and county-level eligibility requirements, including income and the amount of rent owed. Tenants are then placed on a waitlist and will be considered on a first-come-first-served basis. 
When the tenant reaches the front of the list, a social worker examines their funding request. If they are deemed eligible and can provide the required documentation, they will receive a payment covering the rent owed.

There are many issues with this reactive process, and the county is seeking our help in improving it. First, it puts heavy logistical strain on tenants facing eviction, who have to know to contact the helpline or other housing stability organizations to apply for assistance, and must be able to prove their eligibility with documentation.
This requirement excludes all individuals at risk of homelessness due to eviction who do not proactively seek help. For those who apply, the substantial delay between application and receipt of funding means that many tenants have already been evicted by the time they are considered for assistance, or their rental debt has substantially increased. Since eligibility requirements change over time and are not always easily accessible, the current decision system is opaque to those seeking assistance. Finally, rental assistance is not distributed according to who has the most need, but instead to those who know to call the helpline, get through the waitlist in time, and meet the eligibility requirements before being evicted.

\subsection{Related Work}
Previous work has aimed to predict future homelessness to identify those most in need and more efficiently allocate homelessness prevention resources \cite{fowler_2019}. In New York City, re-evaluation of an existing homelessness prevention service resulted in a model based on 15 risk factors that would have outperformed human judgment in selecting at-risk individuals \cite{shinn_2013}. 
In another work focusing on those who were previously homeless, administrative data was used to better predict future homelessness within two years to best match people with programs, though this also disadvantaged other service recipients~\cite{kube_2019}. Because of these tradeoffs, full automation of resource allocation is not recommended: human decision-makers should still be in control.

Work by the California Policy Lab predicted the risk of homelessness among all residents in Los Angeles County~\cite{wachter_predicting_nodate}. Several factors were found to be indicative of future homelessness, including repeated and recent interactions with government agencies or county services and prior mental health crises. Their predictive model achieves 29\% precision for all residents.

\paragraph{Research Gaps.} To our knowledge, no previous work has predicted future homelessness among those who are facing eviction.
he California Policy Lab's work likely suffers from label bias due to the low percentage of homeless individuals utilizing county services (see Section \ref{ssec:limitations}), while our focus on informing a concrete action, rental assistance, allows our system to fit into the existing support infrastructure.

\section{Predictive Approach}

We develop ML models to facilitate need-based prioritization for proactive distribution of rental assistance.
One limitation we encounter with this type of data is that it is conditioned on past rental assistance allocations and offers no insights into the counterfactual outcomes.
As a result, our approach resembles an augmentation of the existing process. This is pertinent not only for the technical implementation but also for the interpretation of our results (e.g., in Section~\ref{ssec:Predictive-models-discover-people}, we concentrate on the outcomes for vulnerable individuals who are not served under the current practice) and the conceptualization of the overall project, which is centered on social impact. and equity.
In Section~\ref{sec:validation}, we outline the transition from this stage to a fully deployed system.

\subsection{Problem Formulation}
\begin{figure}[!htbp]
    \centering
    \includegraphics[width=\columnwidth]{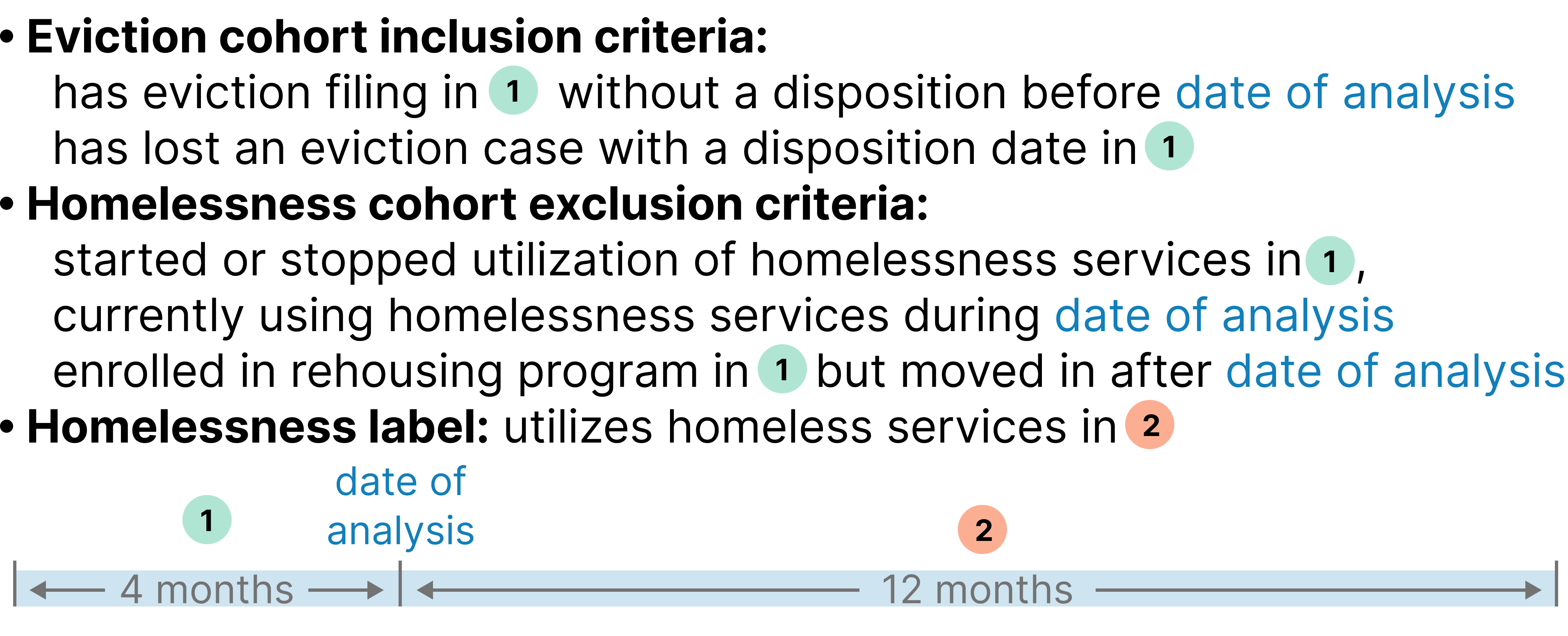}
    \caption{Inclusion/exclusion criteria for homelessness and eviction. Indicators for homelessness in \ACDHS data include clients' interactions with homelessness services, such as staying in a shelter. We also include clients enrolled in rehousing programs that have not moved in as of the prediction date, as being homeless is a prerequisite for enrollment.}
    \label{fig:definitions}
\end{figure}

We formulate our task as a binary classification problem: identifying tenants that will interact with homelessness services within 12 months of the date of prediction. At each prediction date, we consider individuals who have an active eviction filing against them within the past four months and are not currently homeless.\footnote{More details on the demographic composition of this group are provided in Supplementary Materials~\ref{app:Demographics}.} From this cohort, our models identify the 100 individuals with the highest chance of interacting with homelessness services in the next 12 months. We select 100 individuals since this corresponds to the monthly intervention capacity of \ACDHS. Specific inclusion and exclusion criteria are detailed in Figure~\ref{fig:definitions}.

\subsection{Data and Feature Engineering}
\label{ssec:data}
\ACDHS collects and combines administrative data from various county and state-level programs, including information on individuals who have previous eviction filings, homelessness spells, interactions with mental, behavioral, and physical health institutions, address changes, or who have been enrolled in a variety of other \ACDHS and state programs.

For each client at each analysis date, we generate $\sim$7000 features based on the data sources described in Supplementary Materials~\ref{app:Current-Data-Sources}. The features can be classified as follows:

\paragraph{Demographics and Event features.} For each tenant, we include the total number of interactions with each \ACDHS and state program, total number of evictions, and total number of physical and mental health visits. We also include categorical features, such as the particular type of physical or mental health visit and demographic features (race, gender, and age).

\paragraph{Temporal aggregation features.} For each data source, we generate several temporal features to capture the dynamic nature of the process and assess how frequently an individual interacts with state and local services within the last 3 \& 6 months, and 1, 2, 3, 4, and 5 years. For example, using the eviction data, we generate the following features per individual: number of days since most recent eviction; number of evictions in the specified time period; sum, min, max, average rent owed in eviction cases; min, max, average inter-arrival times between evictions.

\subsection{Model Training and Validation} 
\label{ssec:model-training}
We use various supervised classification methods to predict entry into homelessness: logistic regression (LR), decision trees (DT), random forests (RF), Adaboost, Light Gradient Boosting Model (LGBM), and XG boost using Scikit-Learn~\cite{scikit-learn} and the hyperparameter grid specified in Supplementary Materials~\ref{app:param_grid}. To most closely reproduce the context in which our models will be deployed, we use temporal validation~\cite{DBLP:journals/nca/Hoptroff93}, generating different training and evaluation matrices for each analysis date. For example, when evaluating the efficacy of predicting homelessness as of January 2019, we train models on feature label pairs using data up to January 2019, and evaluate them based on how many people become homeless between January 2019 and January 2020 (see Supplementary Materials~\ref{app:Temporal_Validation}).
Temporal validation requires that no information beyond a certain date is used to evaluate an algorithm to avoid data leakage, i.e., predicting the future using data from the future, which can be challenging when combining real-world data from disparate sources that are updated at different frequencies (see Section \ref{ssec:lessons} for a full discussion).

\subsection{Baseline Models} 
\label{ssec:baselines}

We compare the ML models to several simple baselines that either attempt to approximate \ACDHSs current system of allocating rental assistance, or that provide simple improvements on the current status quo. Baselines represent simple heuristics that do not require implementing an ML model (i.e., ranking individuals based on a single attribute) and are, therefore, much easier to deploy and understand.
\begin{enumerate}[noitemsep,label=B\arabic*.,align=left]
    \item \textbf{Previous homelessness.} Prior homelessness is a strong indicator of future homelessness~\cite{glendening_risk_2017}. With this in mind, \ACDHS could prioritize clients by the last date they interacted with any homelessness service. Here, a more recent date would imply a tenant is at higher risk of future homelessness.
    \item \textbf{Baserate.} If \ACDHS were to randomly select individuals to give rental assistance, the precision of the approach would be equal to the proportion of individuals in our cohort who become homeless in the next year, around 2\% during our period of analysis.
    \item \textbf{Earliest OFP.} As an approximation of the current process' first-come-first-serve waitlist, we look for tenants who have been waiting for the longest since an \textit{Order for Possession (OFP)} has been granted, which allows the landlord to evict the tenant.
\end{enumerate}

\noindent Other baselines were omitted from these results due to poor performance (see Supplementary Materials~\ref{app:baselines} for a full list).

\section{Key Findings}
\label{sec:key-findings}
We trained and validated over 5000 model variants to select the 100 individuals with the highest score, i.e. the highest risk of falling into homelessness within the next 12 months. The model selection metrics are measured at top-$k$, where $k=100$.

\subsection{Predictive Models Are More Efficient and Effective than Heuristic Baselines}
\begin{figure}[!t]
    \centering
    \includegraphics[width=0.47\textwidth]{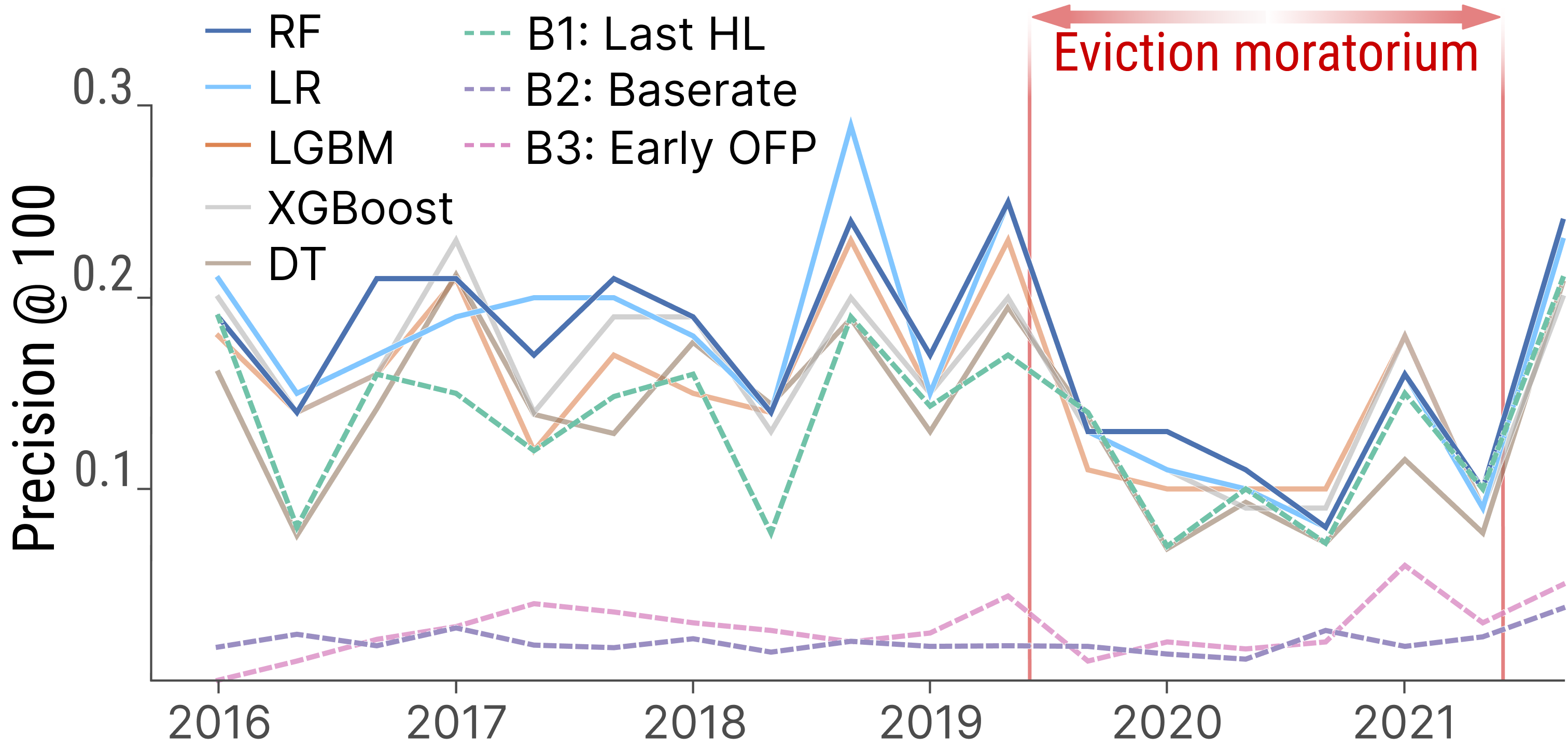}
    \caption{\textit{LR and RF outperform all baselines:} Precision@100 over time shows that out of all baselines, \blhomelessfull performs best, but LR and RF perform better for all splits outside of the moratorium.}
    \label{fig:prec100}
\end{figure}

We evaluate the performance of predictive models based on their ability to improve the efficiency (i.e., ensuring assistance goes to those who are most in need) and effectiveness (i.e., maximizing the reach to people who would otherwise become homeless) of the allocation of rental assistance resources while satisfying the equity constraints. Efficiency and effectiveness are measured by precision@100 and recall@100 respectively.
For each model, we calculate these metrics across all temporal validation splits and explore a broad hyperparameter space for each model type.
Table~\ref{tab:model_performance} summarizes the metrics over time. We primarily focus on the average values of both metrics as we aim to select models that will generalize well to the future. Based on guidance from our partners at \ACDHS, we exclude validation splits that coincide with the COVID-19 eviction moratorium period when measuring average precision\footnote{During this time, there are very few people facing eviction due to the moratorium on evictions during COVID-19 so our cohort becomes very small. See Supplementary Materials~\ref{app:COVID-19-effects}.}. Figure \ref{fig:prec100} shows precision@100 over time for the best hyperparameter configuration for each model type.

We observe that RF and LR perform best and outperform simple heuristic baselines with respect to precision@100, showing a $\sim$20\% improvement over the best heuristic, \blhomeless, and performing 10x better than our approximation of the current allocation process, \blofp.

\begin{table}[!htbp]\textbf
\centering
\begin{tabular}{l l @{\hspace{7pt}} l @{\hspace{7pt}} l @{\hspace{2em}} l @{\hspace{7pt}} l @{\hspace{7pt}} l}
\toprule 
\multirow{2}{*}{Model type} & \multicolumn{3}{c@{\hspace{2em}} }{Precision@100} & \multicolumn{3}{@{\hspace{-4pt}}c}{Recall@100} \\ 
                & Avg & Min & Max & Avg & Min & Max      \\ 
 \midrule

\textbf{RF} & \textbf{0.20} & \textbf{0.14} & 0.25 & \textbf{0.22}  & 0.16 & \textbf{0.34}  \\
LGBM & 0.18  & 0.12 &  0.23 & 0.19 & 0.14 & 0.33  \\
\textbf{LR} & \textbf{0.20}  & \textbf{0.14} &  \textbf{0.29} &\textbf{0.22} & \textbf{0.17} & 0.33  \\
DT & 0.16 & 0.08 & 0.21 &  0.18 &  0.09  &  0.30  \\ 
XGBoost & 0.18 &  0.13 &  0.23  &  0.20 & 0.16  & 0.29    \\ 

\midrule

\textit{\blhomeless: Prev. HL} & 0.15  & 0.08 & 0.21   & 0.17 & 0.09 & 0.30 \\
\blbaseratefull& 0.02   & 0.01 & 0.04   & 0.02 & 0.02 & 0.05 \\ 
\blofpfull & 0.03  & 0.00 & 0.05 & 0.03 & 0.00 & 0.07\\ 

\bottomrule         
\end{tabular}
\caption{\textit{RF and LR outperform:} Precision and recall over time for the model types considered show that our models outperform even the best baseline, \blhomelessfull.}
\label{tab:model_performance}
\end{table}
\subsection{Predictive Models Identify People Who Are Overlooked by the Current Process}
\label{ssec:Predictive-models-discover-people}

The risk in relying on tenants to proactively apply for rental assistance is that the people most in need will not receive it and will end up homeless. 
In each cohort, about 75 people become homeless within a year of their eviction filing. Notably, a majority of these people (50 on average) do not apply for help. This \missedgroup---those who interacted with the homelessness system within 12 months of the prediction date, did not apply for help, and did not receive rental assistance---illustrates the limitations of the current reactive allocation process.

\begin{figure}[!htbp]
        \centering
        \includegraphics[width=\columnwidth]{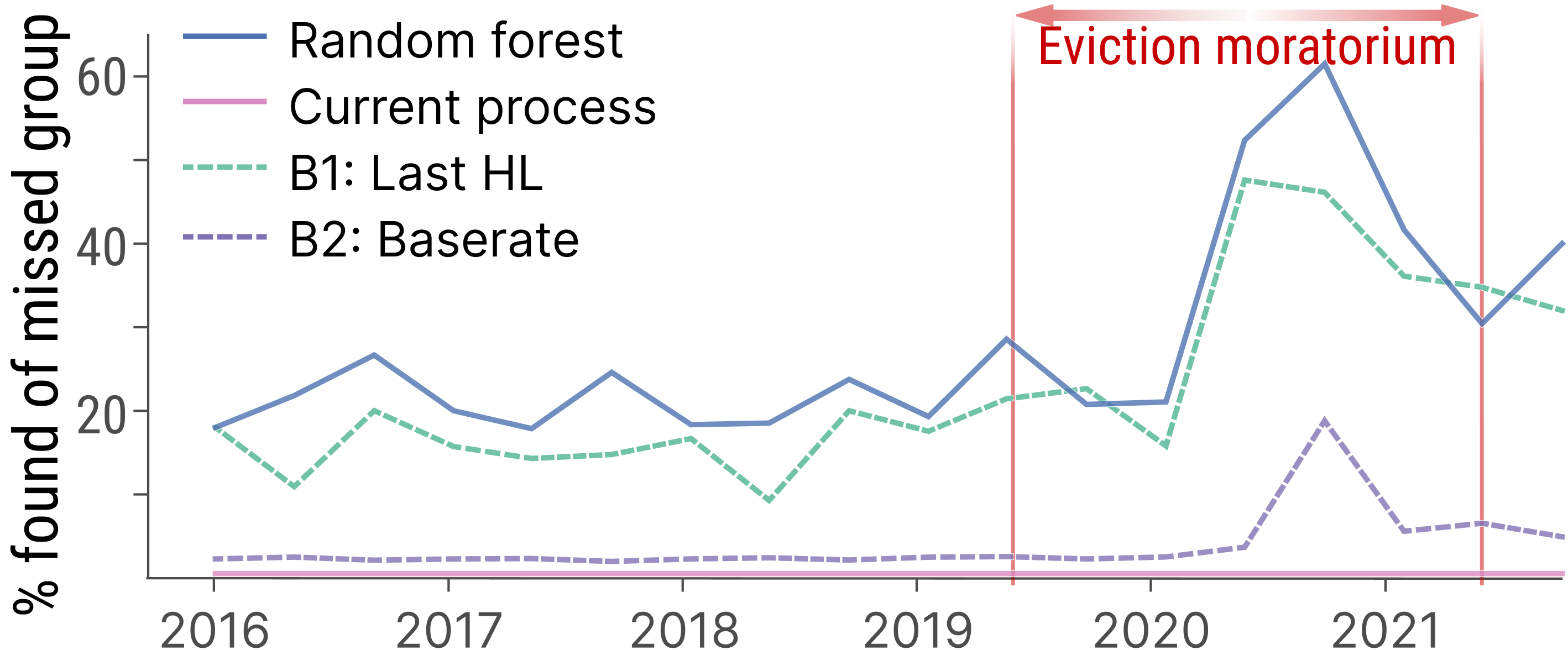}
        \caption{Percentage found of \missedgroup, the individuals who become homeless having not contacted \AlleghenyCounty Link or received rental assistance.
        }
        \label{fig:no_assistance}
\end{figure}

In Figure~\ref{fig:no_assistance}, we examine what percentage of the \missedgroup is found by each model, comparing one of our best ML models (RF) to the baselines \blhomelessfull and \blbaseratefull. \blbaserate finds 4\% of this group on average, meaning that proactive outreach to a list of 100 people from this random approach would find about 4 people every month who would be overlooked by the current rental assistance practice and would end up homeless. 
Both \blhomeless and RF reach a substantially larger group of residents, 23\% for \blhomeless and 28\% for RF on average, potentially providing rental assistance to an additional 10--20 people from the \missedgroup in each cohort every month.

\subsection{Predictive Models Can Promote Fairness and Equity} 
\label{ssec:results-fairness}
Efficiency and effectiveness metrics do not reveal whether allocating resources according to these models would be biased against vulnerable groups \cite{fairMLbook}.
Since we are informing the allocation of a scarce resource, we use \emph{equality of opportunity} \cite{hardt2016equality} as the fairness principle,
which is captured using the true positive rate (TPR) of vulnerable subgroups in the top 100.
We consider the following groups: 
\paragraph{Race.} Black individuals are at higher risk of falling into homelessness in \Allegheny County. Out of those that will become homeless in our validation cohorts, on average, ~60\% were Black, even though Black individuals make up only 14\% of the county's population ~\cite{census_2022}. 
To mitigate these disproportionate impacts, ML models should have a higher TPR compared to white individuals (i.e., serving a higher proportion of Black individuals who are actually at risk of homelessness).
We calculate $ \frac{P(D=1|Y=1, A={black})}{P(D=1|Y=1, A={white})}$ where D, Y, and A represent the decision, outcome, and attribute of interest (e.g., race) 
respectively, finding average recall ratios of 1.34 and 1.14 respectively across temporal splits, indicating that our models meet our fairness considerations for race.  
    
\paragraph{Gender.} For gender, we consider men and women since data on non-binary and transgender individuals is not well-documented in \ACDHS data.
We calculate $\frac{P(D=1|Y=1, A={female})}{P(D=1|Y=1, A={male})}$ for our best models and see that RF and LR models are slightly under-serving women, with average TPR ratios of 0.9 and 0.87 respectively. This issue needs to be considered further in the resource allocation process.

\subsection{Prior Homelessness and Mental Health Crises Contribute to Homelessness Risk}
\label{ssec:feature-importance}

We find that previous use of homelessness services and interactions with mental and behavioral health services are consistently predictive of future homelessness spells across all validation temporal splits.  Most predictive features include:

\paragraph{Previous homelessness service utilization.} The number, duration, and recency of past homelessness spells are highly predictive of future homelessness. Past referrals to homelessness services, emergency shelter utilization, and public housing utilization are identified as highly predictive. Interestingly, the most predictive feature was the number of days since the last homelessness spell (baseline \blhomeless). 

We also compared the characteristics of the top 100 individuals versus the rest of the tenants. We found that for the validation split starting on Sept. 1, 2021, the top 100 individuals were \textit{34x more likely to have been in an emergency shelter} in their lifetime, 31x more likely to have been homeless, and likely to have spent 28x as many days in homelessness compared to other tenants. 

\paragraph{Mental and behavioral health service interaction.} Mental and behavioral health events and related service utilization are highly predictive of future homelessness. In the best performing RF, the number of days since the last mental health or behavioral health crisis event, the duration of mental health service utilization, the number of times one used mental health services, and the number of mental health crisis events were highly predictive features. 

Compared to the rest of the cohort, those in the top 100 were \textit{100x more likely to have had a mental health crisis event} in the last three years and 34x more likely to have had one in their lifetime, likely to have had 28x as many days utilizing mental health services, and likely to have had 24x more behavioral health events in their lifetime.

\subsection{First-Time Homelessness is Harder to Predict}
About half of the people in a cohort who become homeless have not previously been homeless, but we find that our models rely on features about previous homelessness spells, resulting in lower recall for people experiencing first-time homelessness.
Our models perform substantially better on people with a history of homelessness: RF's average recall (excluding the eviction moratorium) is 55\% for people who have experienced previous homelessness but only 4\% for people experiencing first-time homelessness. We are exploring  1) building separate models, and 2) adding additional data sources to better predict future homelessness based on whether or not the person has previously been homeless.

Interestingly, the group who receives rental assistance through the existing practice appears to mostly be people without a history of homelessness: for the validation split starting on Sept. 1, 2021, over 70\% of people in our cohort who apply for help and over 80\% of people who receive rental assistance have not been homeless before.

\subsection{On False Positives: Models Identify Vulnerable Individuals}

Out of those facing eviction in \Allegheny County, about 2\% end up interacting with homelessness services within 12 months.
With such heavy class imbalance, it is inevitable that we have false positives in our top-k, i.e. recommend giving rental assistance to those who will not become homeless within 12 months. Who are these false positives? Are they vulnerable to other adverse outcomes, including homelessness beyond 12 months or other crises?

\begin{figure}[!htbp]
    \centering
    \includegraphics[width=.92\columnwidth]{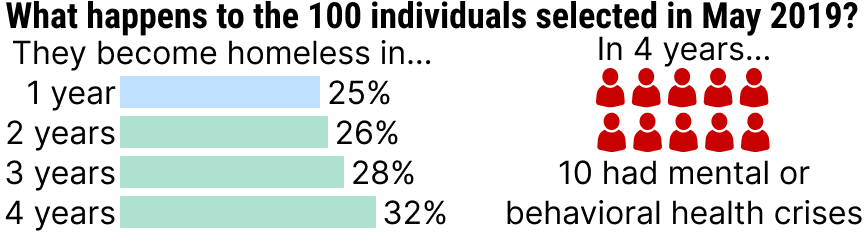}
    \caption{\textit{False positives are still vulnerable:} among these tenants, we see that homeless service utilization and mental or behavioral health crises are common beyond 1 year.
    }
    \label{fig:wrong-predictions}
\end{figure}

In Figure~\ref{fig:wrong-predictions}, we investigate the top 100 picked by the RF model as of May 1, 2019 as it provides a sufficient temporal gap from the present to observe long-term outcomes. 
We see that even though the predictions are ``wrong'' about who will interact with homeless services within 12 months, they highlight people vulnerable to homelessness and other adverse outcomes and in need of assistance.

\section{Field Validation}
\label{sec:validation}

Before deployment, we validate our results in two consecutive stages:
first, shadow mode deployment (SMD) alongside the current process to validate our solution on live data, and second, a planned randomized control trial (RCT) to compare our proposed solution to the status quo and assess the effectiveness of the treatment (rental assistance).

\subsection{Shadow Mode Deployment (SMD)}
\label{ssec:SMD}
To validate our model against real-time data with no chance of data leakage (see Supplementary Materials~\ref{app:Current-Data-Sources}), we use our model to make predictions in real time, while the decision-making process continues to use the current system.
We used the best-performing model that met our equity goals as of September 1, 2022, and trained it on all data available at the time to produce a list of 100 individuals at risk of falling into homelessness within 12 months.
We then compared this list to the current rental assistance distribution process.

Between September 2022 and August 2023, 22 of the 100 individuals on the predicted list made use of homelessness services, confirming our model's \precision of $\sim$0.20. There is little overlap between individuals targeted by the current approach and those that would have been targeted using our predictive model: among our list of 100 individuals, only 12 received rental assistance.%
\footnote{Of those 12, 4 used homelessness services within 12 months. Assuming rental assistance is effective, our SMD \precision of 0.22 represents a lower bound, but this will have to be tested in our RCT (see Section \ref{ssec:field-trial}).}
Importantly, our model would have resulted in proactive assistance to 17 people who are \textit{missed} by the current system, i.e. to those who did not reach out for assistance, did not receive rental assistance, and ended up falling into homelessness within a year.

\subsection{Randomized Control Trial (RCT)}
\label{ssec:field-trial}
To accurately assess the \textit{efficiency} of both the current and the ML-based process, we must compare rates of homelessness among individuals who received rental assistance to the counterfactual case where such individuals would not have received assistance. Similarly, to assess the \textit{effectiveness} of rental assistance in reducing homelessness, we must compare rates of homelessness between those that did and did not receive assistance, all else being equal. Such causal identification is not possible through historical observational data.

To compare the impact of the current and proposed allocation process, we are planning a randomized control trial (RCT). In principle, an RCT should:
\begin{enumerate}[topsep=.1pt,itemsep=-1ex,partopsep=1ex,parsep=1ex]
    \item select two sets of eligible candidates, one using the current approach and one using the algorithmic approach,
    \item randomly assign $k\%$ individuals to a treatment and $(1-k)\%$ to a control group for each candidate set.
\end{enumerate}
To compare the \textit{efficiency} of our new system to that of the current system, it would then be possible to compare homelessness rates among individuals selected by both approaches who were assigned to the control group (those who did not receive rental assistance), i.e. $P(Y = 1| D_{ML} = 0) - P(Y = 1| D_C = 0)$, where $D_{ML} = 0$ and $D_C = 0$ denotes random assignment to the control group in the algorithmic ($D_{ML}$) and current list ($D_C$). Similarly, to assess the \textit{effectiveness} of rental assistance at reducing homelessness, it would be possible to compare rates of homelessness among those in the treatment to rates of homelessness among those assigned to the control group (i.e. $P(Y = 1| D_{ML} = 1) - P(Y = 1| D_{ML} = 0)$ and  $P(Y = 1| D_C = 1) - P(Y = 1| D_C = 0)$).\footnote{Note that the effectiveness of the treatment may vary between people targeted by the current process and those targeted by the ML-based solution, and even different ranges of risk scores between the two. These potential differences should also be assessed.}

There are ethical challenges involved with randomizing the allocation of rental assistance. In particular, this RCT would entail not providing assistance to eligible people who applied for funding and who would have received it in the absence of the RCT. Instead, we could rely on a \textit{quasi-random} assignment of individuals to the control group, as some happen to call the helpline on days when the waitlist is full or funding is not available ~\cite[see][]{evans_2016}. This would allow us to construct a counterfactual group of eligible individuals who do not receive funding without interfering with the current process. We are currently discussing the RCT design options with \ACDHS and plan to finalize them soon.

\section{Limitations}
\label{ssec:limitations}
\paragraph{Eligibility Constraints.}
We decided not to consider eligibility requirements for the following reasons. There are multiple sources of eligibility requirements---those tied to funding sources and the internal ones set by \ACDHS---which change over time and are not well-documented. Not only is it impossible to encode historical eligibility requirements given the lack of data, but we also want to alert \ACDHS to individuals with the most need, regardless of their eligibility.

\paragraph{Label Bias.}
Our labels are likely biased due to \textit{outcome measurement errors}, which are common in applying predictive modeling approaches to observational data ~\cite{Coston2020Counterfactual,Jacobs2021Measurement}.
Utilization of homelessness services is just a \textit{proxy}, as not all who fall into homelessness use these services. Individuals who are couch surfing, sleeping in their cars, or unsheltered on the street will not be captured by our outcome definition (label).

In \Allegheny County, the large majority of homeless individuals use homelessness services, such as emergency shelters,\footnote{\ACDHSs most recent point in time count, where workers manually count all homeless individuals in the county on one day, found that 83\% of homeless individuals were staying in shelters that day~\cite{pit_count}.} at some point during their homelessness spell, particularly in winter. Studies conducted in warmer climates, however, are likely to systematically miss those who never make use of any homelessness services~\cite[e.g.][]{wachter_predicting_nodate}. By selecting 12 months as our label period, we guarantee that our label captures shelter use during the winter months.

\section{Lessons Learned}
\label{ssec:lessons}
Throughout this process, we've learned several lessons about how to most effectively use AI methods in a real-world, resource-constrained context, many of which generalize beyond the allocation of rental assistance.

\paragraph{Scoping: the underrated first step.}
What problem actually \textit{needs} solving? What will \textit{actually} be used in practice? While these questions may seem like the obvious first step, a common question instead asked by many is ``what are interesting research problems we can explore with this data?,'' regardless of their real-world impact. We spent months scoping~\cite{scopingguide} the project with domain experts focusing on societal goals and actions we can inform until \ACDHSs specific needs were clear. Only then did we formulate that need as a modeling problem and explore ML-based approaches.

\paragraph{Data leakage: the secret deceptor.}
What data will be available at the prediction date, when decisions are made? 
Real-world data is messy: columns are populated on different dates and updated at different intervals. Unwittingly using data that was updated after the prediction date can artificially inflate performance but data leakage can be difficult to detect. Early on, we discovered that the feature denoting if a resident's age was imputed was a strong indicator of future homelessness. Upon further investigation, we found that the source containing age was updated in place, meaning that if a person had interacted with a program that recorded age \textit{after} the prediction date, it would be reflected in our feature.
It became a signal for homelessness because they had interacted with social services after the prediction date. Unknowingly, we had run into data leakage.

\paragraph{Field trials: the reality check for ML's social impact.}
Often, a field trial is necessary to construct counterfactual outcomes and compare the effectiveness and efficiency of the proposed solution to that of the current process. However, randomization can be ethically challenging as it involves not providing help to otherwise eligible individuals in need. In such cases, researchers should aim to construct counterfactual groups in a manner that minimizes interference with the current allocation process if feasible.

\paragraph{Evaluation for real-world impact.}
Metric selection and validation techniques require careful consideration of the real-world problem. Standard evaluation methods, such as k-fold CV and AUC, may not accurately capture the success of a model for the given task. We started by defining our specific goals, e.g. efficiency, effectiveness, and equity, and then found the appropriate metric for each goal. In resource-constrained policy problems like ours, precision@$k$ represents an appropriate efficiency metric, while AUC or F1 scores are less meaningful.

\paragraph{Communicating to policymakers.}
Policymakers  may not have the technical expertise to decipher ML models or metrics. We need to assist them in making informed decisions, e.g. by providing intuitive explanations of metrics and a palette of model options with their associated policy tradeoffs.
While this can be time-consuming, the alternative can lead to model misuse or output misinterpretation.

\section{Conclusion}
We have shown that predictive modeling can improve upon the prioritization of rental assistance to tenants facing eviction in \AlleghenyCounty to reduce the rate of entry into homelessness. Our novel approach has four main contributions, providing (a) need-based prioritization of rental assistance (b) in a proactive manner, which is at least 20\% more effective than simpler baselines while being equitable. We (c) validated our models by deploying them as a shadow model and are designing an RCT that is being discussed with \ACDHS. We also (d) included some lessons learned that other AI researchers can use to ethically design predictive support tools. After the RCT, this process is expected to be used in practice, improving the effectiveness of allocation of rental assistance and equity in outcomes in \AlleghenyCounty.

\section*{Acknowledgements}
 This work was started as part of the 2022 Data Science for Social Good (DSSG) Fellowship at Carnegie Mellon University and partly funded by a grant from the Richard King Mellon Foundation. We thank Adolfo De Unanue for his valuable input during DSSG. We also thank our partners at ACDHS, particularly Rachel Rue and Justine Galbraith, for their help throughout this project.

\bibliography{ref}

\newpage
\appendix
\onecolumn

\section{Data Sources}
\label{app:Current-Data-Sources}

Table~\ref{tab:DataSources} shows the data sources that were used for feature generation. Data from the court system on eviction was linked using a unique identifier to other county data, including demographics, enrollment in county/state programs, housing and homelessness services, and mental/behavioral health interactions. The dataset contains interactions between January 2012 and August 2023.

\begin{table}[!htbp]
\centering
\begin{tabularx}{\textwidth}{llX}
\toprule
\textbf{Data Type} & \textbf{Info Entry Date} & \textbf{Information} \\
\midrule \midrule

\multirow{3}{2.5cm}{\textbf{Demographics}} & & \\
& most recent interaction & \tablebullet gender, birthdate of individual, and (frequency of) address changes \\

&&\\\midrule
\multirow{5}{*}{\textbf{Evictions}} & & \\
& filing date & \tablebullet dollar amount claimed to be owed according to landlord \\
& hearing date & \tablebullet who won the case (tenant or landlord)\\
& & \tablebullet how much tenant owes landlord \\
& OFP date & \tablebullet whether an OFP has been filed \\

&&\\\midrule
\multirow{4}{2cm}{\textbf{Program Interactions}} & & \\
& enrollment date & \tablebullet program type: i.e. Medicaid, Food Assistance, Homeless Shelter, Medical Assistance Transportation, or other similar programs offered by \ACDHS \\
& termination date & \tablebullet when client is no longer enrolled in the program\\

&&\\\midrule
\multirow{5}{2cm}{\textbf{Public Housing}} & & \\
& enrollment date & \tablebullet housing service type: i.e. Section 8 Voucher, Rapid Rehousing, or similar \\
& move-in date & \tablebullet when and where client was rehoused (can be long after enrollment date)\\
& address change date & \tablebullet when and where client moves to a new location \\

&&\\\midrule
\multirow{4}{2cm}{\textbf{Mental \& Behavioral Health}} & & \\
& interaction start date & \tablebullet interaction type: i.e. walk-in, crisis, or hospital stay\\
& interaction end date & \tablebullet when person left (only relevant for multi-day stays) \\
& diagnosis date & \tablebullet type of official diagnosis: i.e major depression, bipolar disorder, etc\\

&&\\\midrule
\multirow{4}{2cm}{\textbf{Physical Health (ER)}} & & \\
& interaction start date & \tablebullet when person visited the ER\\
& interaction end date & \tablebullet when person left (only relevant for multi-day stays) \\

&&\\\midrule
\multirow{4}{2cm}{\textbf{Children, Youth, and Families (CYF)}} & & \\
& interaction start date & \tablebullet interaction type: i.e. child moved to foster care or group home\\
& interaction end date & \tablebullet when the child moved to a different service or ``aged out'' (turned 18) \\

&&\\\bottomrule
\end{tabularx}
\caption{Sources of data used for feature generation, as well as the dates at which we consider each piece of information to be known for temporal validation. }
\label{tab:DataSources}
\end{table}

\paragraph{Mitigating Data Leakage.} Since we are using temporal validation, we need to ensure that, if we are evaluating an algorithm with data known up to a certain date, we do not use any information that was not known up to that date. Otherwise, we run the risk that the ''leakage`` of information from the future affects past results. For example, if we were to train a model on data up to January 1 2019 and a client had an eviction in December 2018 but an OFP in February, we must make sure we do not use any information about that individual's OFP in our training data.
At first, this may not seem like a difficult task, but it can prove tricky with real-world, messy data. For each column given to us by \ACDHS, when did they know that data by? Do they update that data daily, weekly, or even monthly? Considering these questions is crucial to ensure that our models do not appear to perform better than they would when actually deployed, in case we were inadvertently using information only known in the future. For this reason, we not only explain the type of information provided by \ACDHS in Table~\ref{tab:DataSources}, but also specify which date we know that information by.

As shown in Table~\ref{tab:DataSources}, most data is associated with a specific interaction, allowing us to generate a temporal history for each client.
However, this is not the case for demographic information, as this is continuously updated without keeping track of old entries.
Consequently, the dataset only reveals what demographic information was known about an individual at the time of the most recent interaction.
Certain demographic information, such as race, are more likely to be \texttt{null} for individuals with few interactions with \ACDHS.
Using this information is problematic for temporal validation since it contains important information from the future (i.e., whether an individual had many interactions with \ACDHS up to the last day in the dataset.
This, in turn, could falsify the performance evaluation results as individuals falling into homelessness interact more with \ACDHS, on average.
To avoid this type of data leakage, we made sure to only use demographic information that is collected during every interaction.

\section{Demographic Composition of Cohort}
\label{app:Demographics}

Table~\ref{tab:demographics} describes the demographic composition of the last pre-pandemic cohort as of January 1, 2019. In this cohort (as in others), Women and African Americans are disproportionately at risk of facing eviction, and, conditional on facing eviction, are at higher risk of future homelessness. Fourteen percent of the population in \Allegheny County is African American~\cite{census_2022}, yet this share jumps to 55\% among those facing eviction and to 59\% among those facing eviction that fall into homelessness the following year. Those with a history of homelessness are also more likely to reenter homelessness in the future: while 6\% of individuals in the cohort had been homeless in the past, this share jumps to 37\% among those who end up in homelessness in the following year.

\begin{table}[!htbp]
\centering
\begin{tabular}{lccc}
\hline
\multicolumn{1}{|l|}{\textbf{\textbf{Characteristics}}} & \multicolumn{1}{c|}{\textbf{\textbf{Total}}} & \multicolumn{2}{c|}{\textbf{Becomes homeless}} \\
\multicolumn{1}{|l|}{}                  & \multicolumn{1}{c|}{}       & \textit{No}             & \multicolumn{1}{c|}{\textit{Yes}} \\ \hline
\multicolumn{1}{|l|}{Female}            & \multicolumn{1}{c|}{56.5\%} & 56.4\%                  & \multicolumn{1}{c|}{61.4\%}       \\
\multicolumn{1}{|l|}{African American}  & \multicolumn{1}{c|}{55.1\%} & 55.0\%                  & \multicolumn{1}{c|}{58.6\%}       \\
\multicolumn{1}{|l|}{Has been homeless} & \multicolumn{1}{c|}{6.0\%}  & 5.4\%                   & \multicolumn{1}{c|}{37.1\%}       \\ \hline
                                        & \textit{(N \!=\! 4036)}     & \textit{(N \!=\! 3966)} & \textit{(N \!=\! 70)}            
\vspace{-2mm}
\end{tabular}
\caption{Demographic composition of cohort as of January 1, 2019 across key demographic groups of interest and label outcomes.}
\label{tab:demographics}
\end{table}

\section{Parameter Grid}
\label{app:param_grid}

Table~\ref{tab:hyperparams} shows the different parameter values that were used for the models. Models with each possible combination of these hyperparameters were trained and tested for each date of analysis.

\begin{table}[!htbp]
    \centering
    \begin{tabular}{lrl}
    \toprule
    \textbf{Model Name} & \textbf{Parameter} & \textbf{Values} \\
    \midrule\midrule
    
    \multirow{4}{2cm}{\model{Logistic Regression}} & &\\
    & \param{C} & \texttt{0.001, 0.01, 0.1, 1} \\ 
    & \param{Penalty} & \texttt{L1, L2} \\
    &&\\\midrule
       
    \multirow{4}{2cm}{\model{Decision Tree}} & & \\
    & \param{Max depth} & \texttt{1, 2, 5, 10, no limit}\\
    & \param{Min samples split} &  \texttt{2, 10}\\
    &&\\\midrule

    \multirow{6}{2cm}{\model{Random Forest}} &&\\
    & \param{Number of estimators} & \texttt{1000, 5000, 10000}\\
    & \param{Max depth} & \texttt{5, 10, 25, 50}\\
    & \param{Min samples split} & \texttt{10, 100}\\
    & \param{Min samples leaf} & \texttt{10, 100}\\
    &&\\\midrule
    
    \multirow{6}{2cm}{\model{Light GBM}} &&\\
    & \param{Boosting type} & \texttt{dart}\\
    & \param{Number of estimators} & \texttt{100, 300, 500}\\
    & \param{\# leaves} & \texttt{31} \\
    & \param{Max depth} & \texttt{10, 100} \\
    &&\\\midrule
    
    \multirow{6}{2cm}{\model{XG Boost}} &&\\
    & \param{Booster} & \texttt{gbtree}\\
    & \param{Learning rate} & \texttt{0.01, 0.1}\\
    & \param{Number of estimators} & \texttt{100, 300}\\
    & \param{Max depth} & \texttt{5, 10, 40}\\
    &&\\\midrule
    
    \end{tabular}
    \caption{Grid search parameters for model selection}
    \label{tab:hyperparams}
\end{table}

\section{Temporal Validation}
\label{app:Temporal_Validation}

\begin{figure}[!htbp]
        \centering
        \includegraphics[width=0.9\textwidth]{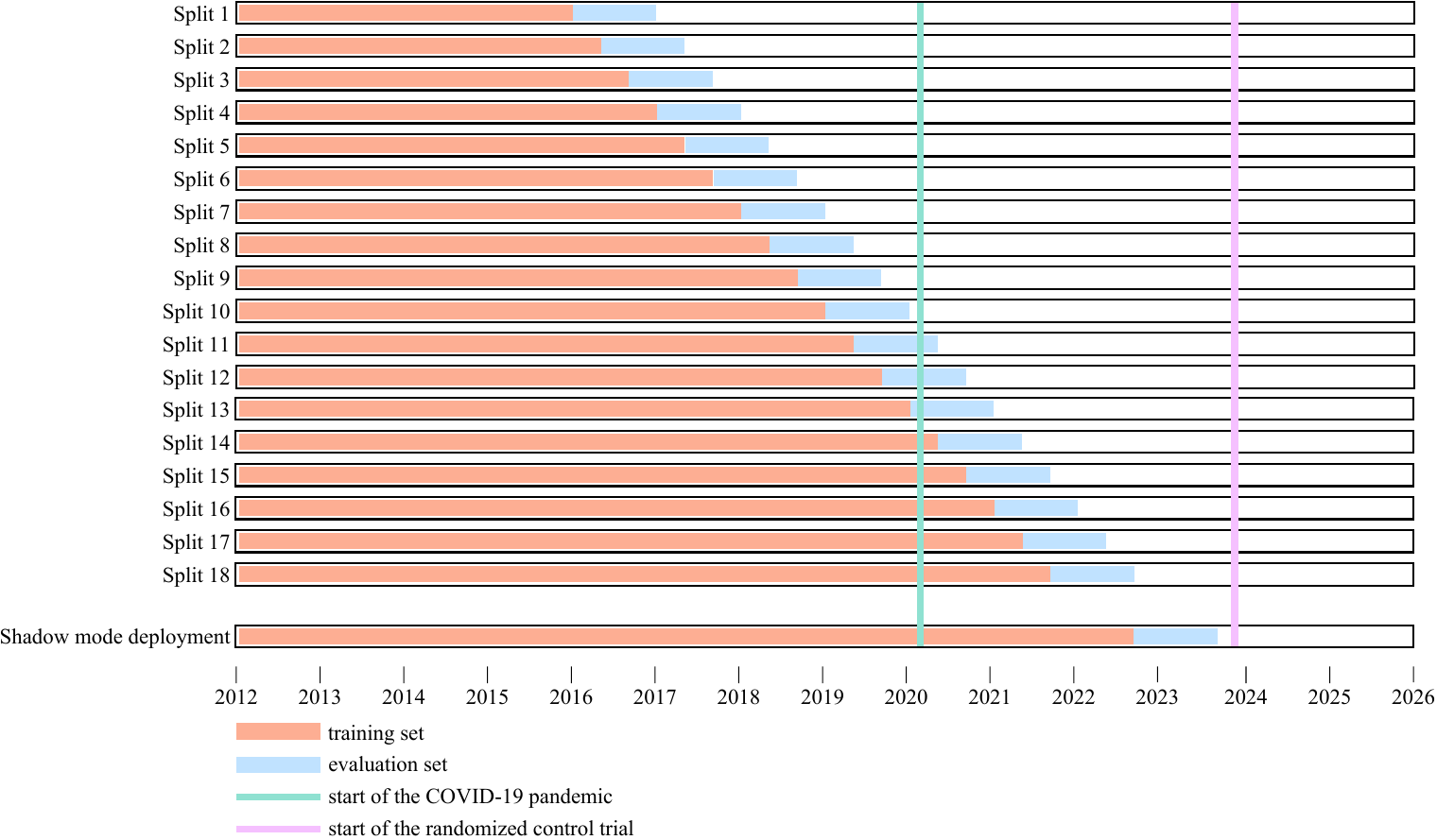}
        \caption{Temporal validation}
        \label{fig:temporal-validation}
\end{figure}

Figure~\ref{fig:temporal-validation} visualizes the temporal validation splits.
The results of splits 1 -- 18 are reported inSection~\ref{sec:key-findings}.
The results of the shadow mode deployment are reported in Section~\ref{ssec:SMD}.

\begin{figure}[!htbp]
        \centering
        \includegraphics[width=0.7\textwidth]{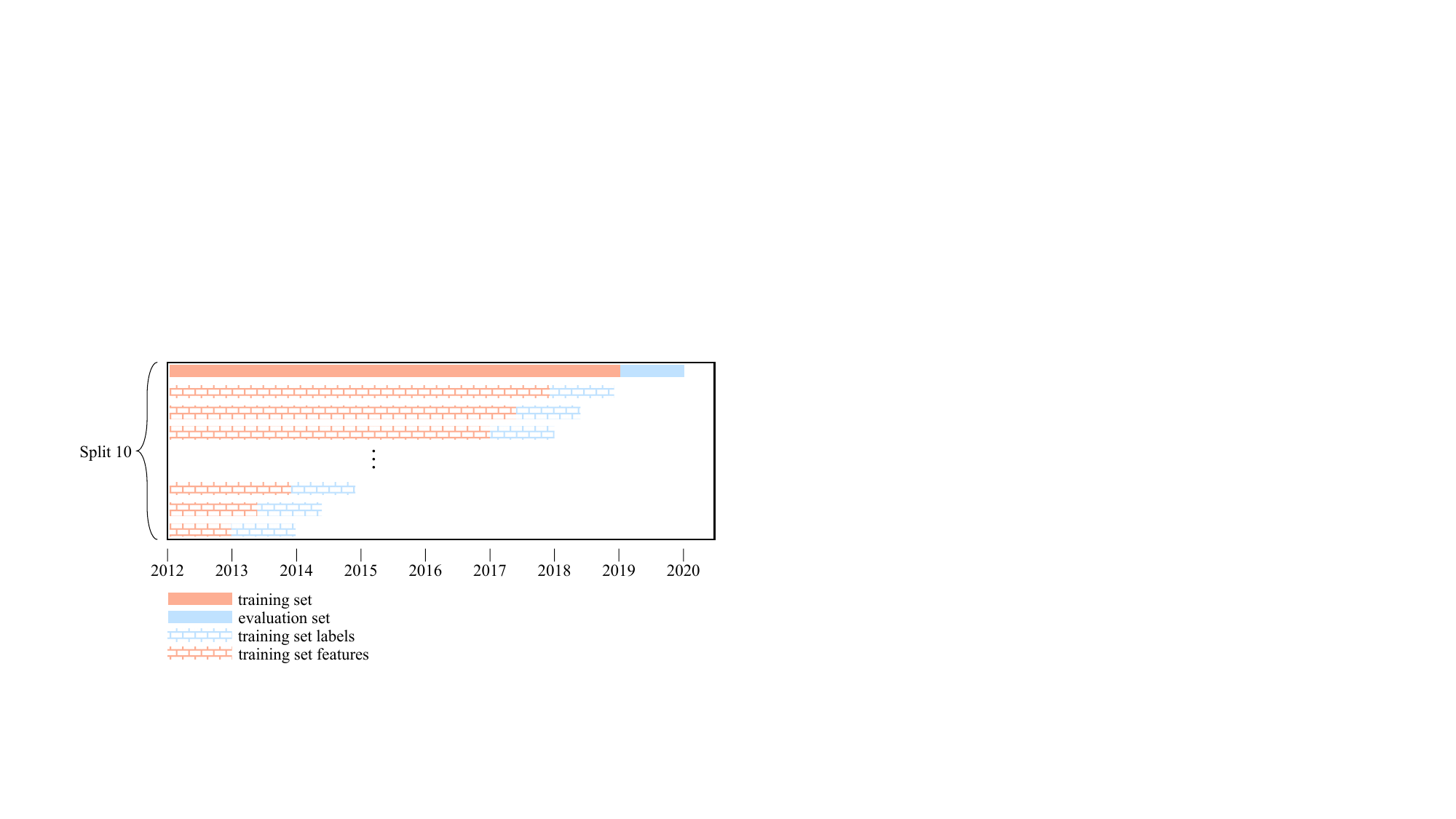}
        \caption{Feature label pair generation in one temporal validation split (here we represent split 10 as an example)}
        \label{fig:temporal-validation-feature-label-pairs}
\end{figure}

For each temporal split, we generate feature label pairs that span the entire timespan of the training set.
As an example, we visualize this for the split 10 in Figure~\ref{fig:temporal-validation-feature-label-pairs}.
The split 10 of the temporal validation corresponds to the model development as of January 1 2019. Thus, the most recent label timespan considered in the training data matrix is between January 1 2018 and January 1 2019.
However, individuals who only made use of homelessness services before January 1 2018 are not labeled positively in this label timespan.
Therefore, we additionally consider label timespans of 12 months, going back in 3 months intervals until January 1 2013.
This is needed to exploit the data available in the training set while respecting the temporal flow of events in the past.

\section{Additional Baselines}
\label{app:baselines}
In addition to the baselines mentioned in Section~\ref{ssec:baselines}, we also tried a few others. These were omitted from the paper due to their poor performance.
\begin{enumerate}[label=B\arabic*.]
    \setcounter{enumi}{3}
    \item \textbf{Age at first interaction.} This baseline sorts by the age at which the individual first was enrolled in an \ACDHS program. The younger the individual at their first interaction, the more likely they are to fall into homelessness.
    \item \textbf{Age at first adult interaction.} Some individuals are involved in child welfare or foster care services from a young age. This baseline extends B4 by only considering \ACDHS program involvement once the individual is an adult (18 years of age).
    \item \textbf{Days since current filing.} Similar to \textit{B1: Current process}, this baseline instead sorts individuals by the date of their current eviction filing (not their OFP date), with earlier dates being considered as more likely to fall into homelessness.
    \item \textbf{Days since last program involvement.} This baseline assumes that individuals who recently interacted with non-homelessness \ACDHS services are more vulnerable, and therefore more likely to fall into homelessness.
    \item \textbf{Number of distinct programs.} This baseline sorts individuals by the number of distinct \ACDHS programs they have been involved in throughout their lifetime, with more programs indicating that the individual is more vulnerable and therefore more likely to fall into homelessness.
    \item \textbf{Number of program involvement spells.} Since an individual can be enrolled in the same \ACDHS program multiple times throughout their lifetime, this baseline extends B9 by considering the distinct number of times an individual has been involved in any \ACDHS program, with more involvement indicating the individual is more likely to fall into homelessness.
    \item \textbf{Total days in program involvement.} Similar to the previous two baselines, this baseline sorts individuals by the total number of days they have been involved in any \ACDHS service, with more involvement indicating the individual is more likely to fall into homelessness.
\end{enumerate}

\begin{figure}[!htbp]
    \centering
    \includegraphics[width=0.6\textwidth]{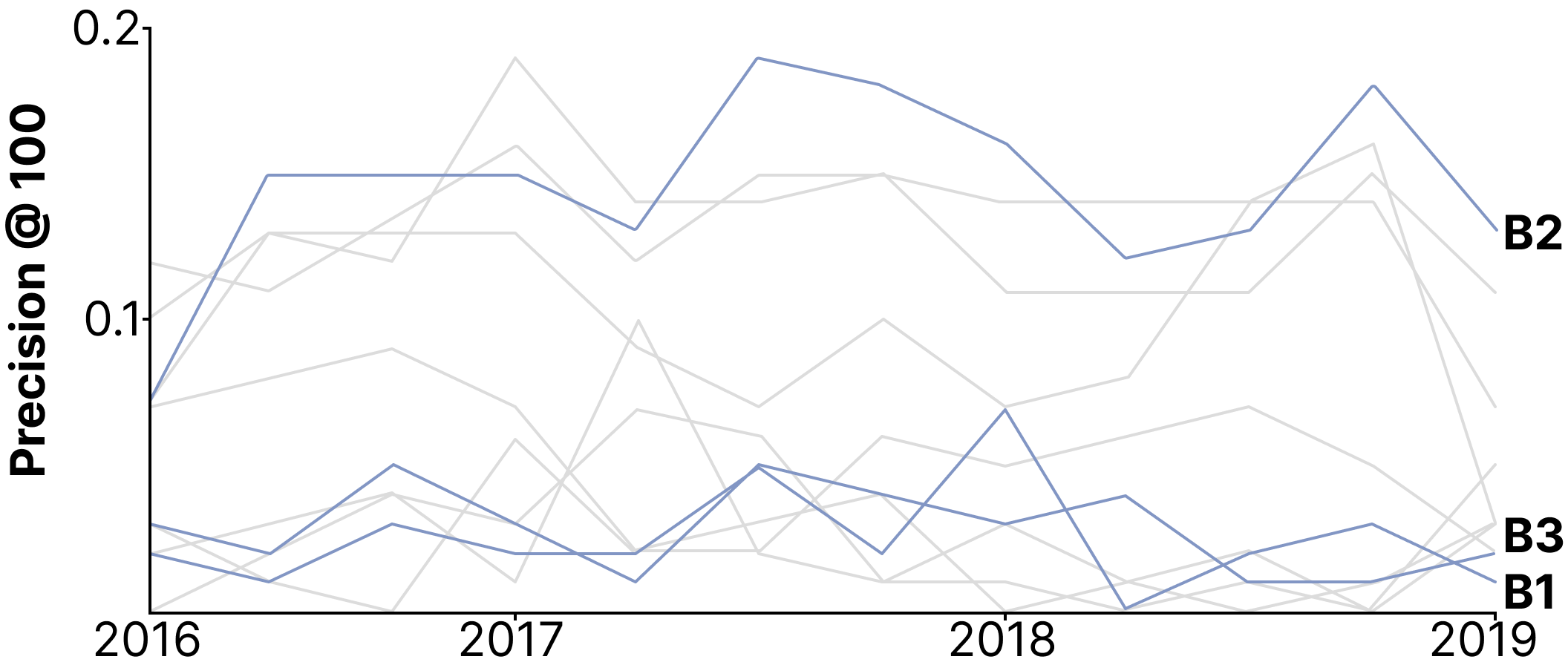}
    \caption{Performance of all attempted baselines: the grey lines showcase the performance of B4 --- B10 which were omitted from the main paper results since they all perform less well than \textit{B2: Previous Homelessness}.}
    \label{fig:baselines_performance}
\end{figure}

Figure \ref{fig:baselines_performance} shows how these baselines perform compared to our selected baselines B1 --- B3. We see that generally, \textit{B2: Previous Homelessness} performs better than other baselines. Though \textit{B1: Current Process} and \textit{B3: Baserate} also do not perform well, they were included in the main results as B1 most closely emulates \ACDHSs current process and B3 shows how well random allocation would perform. 

\section{Effect of the COVID-19 pandemic on cohort size and positive label prevalence}
\label{app:COVID-19-effects}

\begin{figure}
    \centering
    \includegraphics[width=0.6\linewidth]{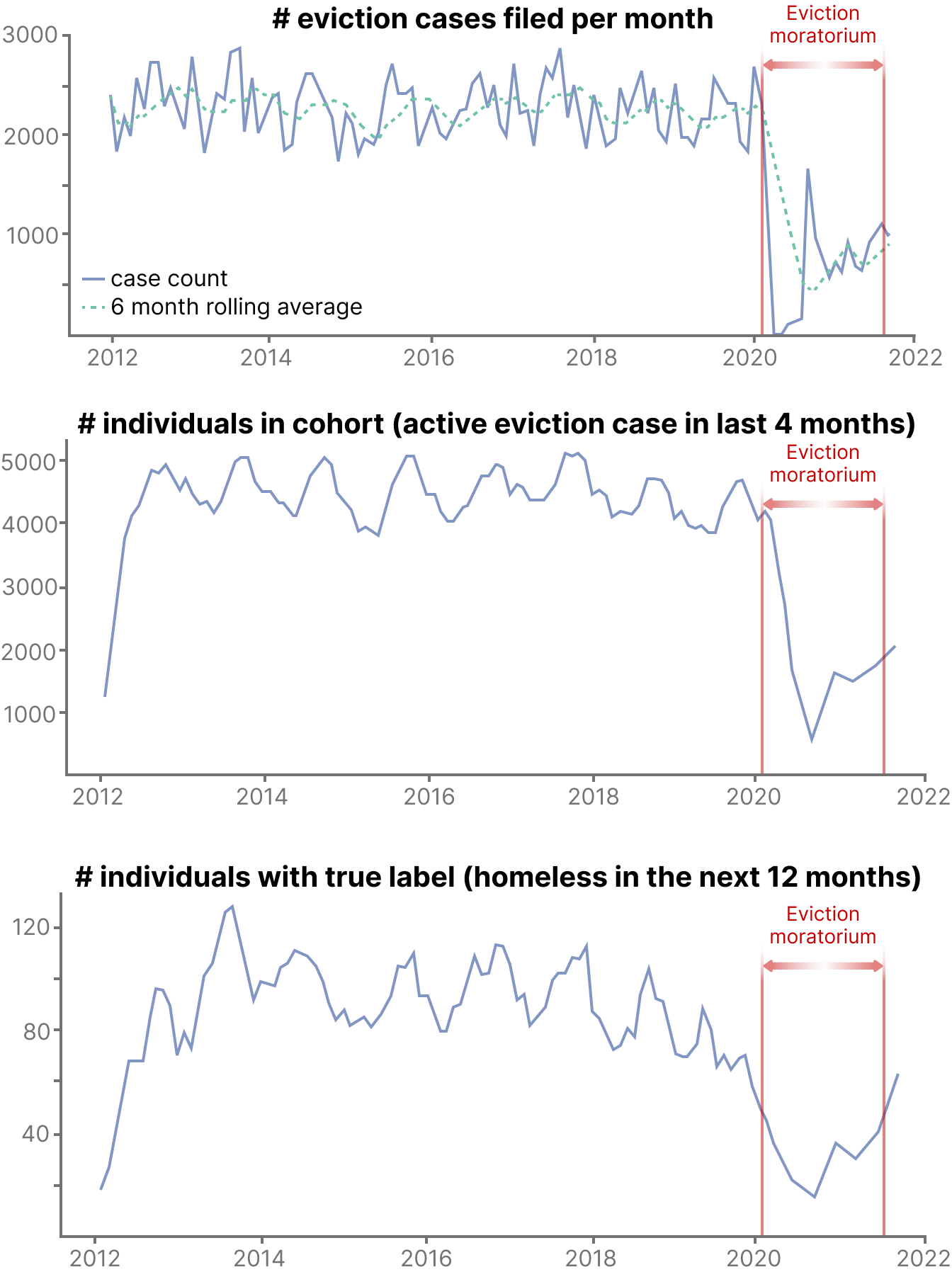}
    \caption{(a) Number of eviction cases filed per month, (b) number of individuals in the cohort as of a particular date of analysis, and (c) number of individuals with that become homeless in the next 12 months as of a date of analysis.}
    \label{fig:pre_post_pandemic_data}
\end{figure}

Figure~\ref{fig:pre_post_pandemic_data}a shows that, due to the eviction moratorium, the number of cases drops drastically in the year 2020.
However, after the moratorium, the number of eviction cases rises almost to pre COVID-19 pandemic level in the year 2022.
As can be seen in Figure~\ref{fig:pre_post_pandemic_data}b, the cohort size also changes accordingly in this time period -- as does the number of positive labels in those cohorts, see Figure~\ref{fig:pre_post_pandemic_data}c.

\section{Field Trial Schematic Design}
\label{app:field-trial}
\begin{figure}[!htbp]
        \centering
        \includegraphics[width=0.52\textwidth]{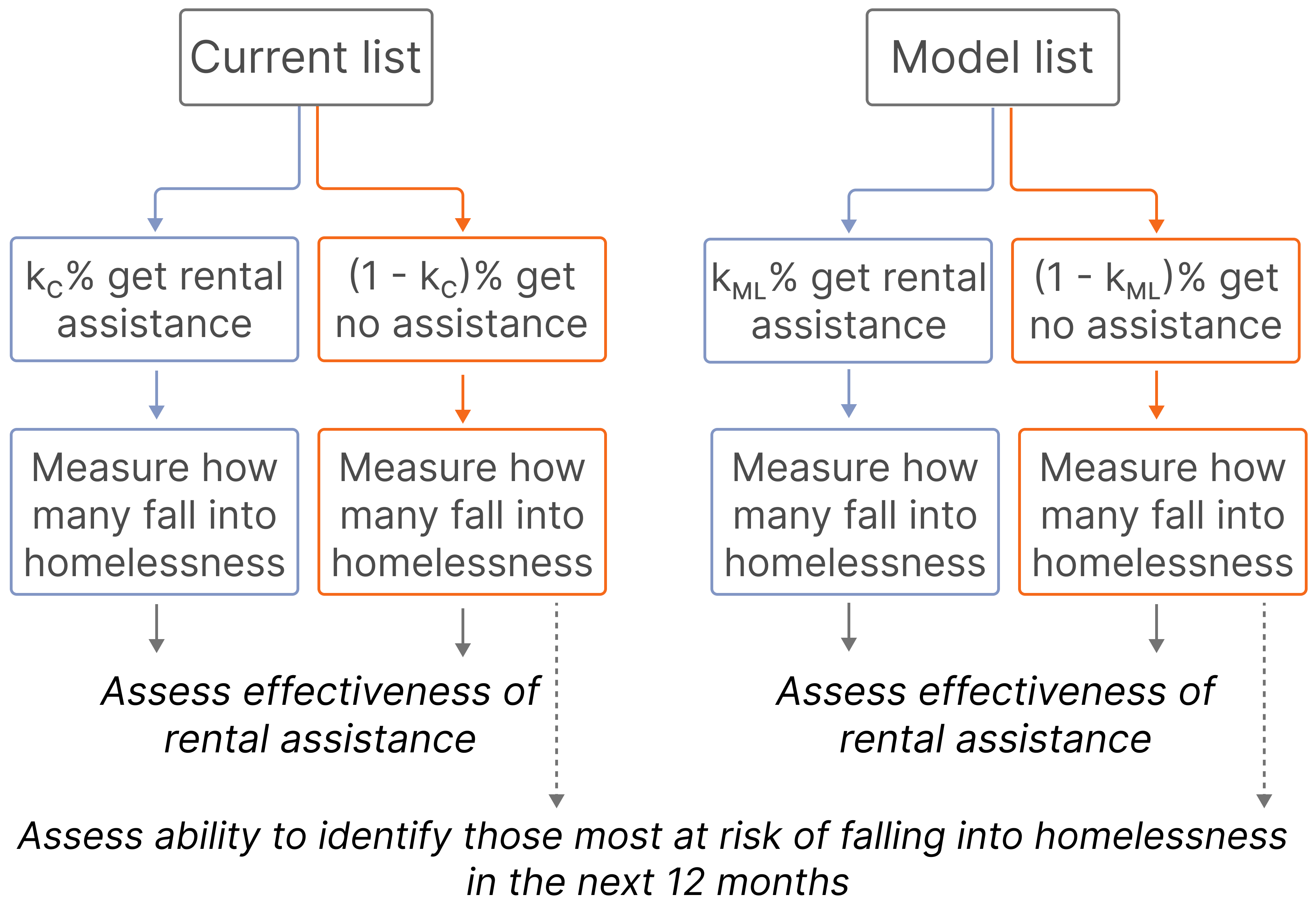}
        \caption{Schematic drawing of RCT design}
        \label{fig:field-trial}
\end{figure}

\end{document}